\documentclass[twocolumn,preprintnumbers,amsmath,amssymb,nofootinbib
,superscriptaddress]{revtex4}
\usepackage{hyperref}
\usepackage{graphicx}
\usepackage{color}
\usepackage{xcolor}
\newcommand{\nonumbo}{\nonumber \\ && }
\newcommand{\backo}{\!\!\!\!\!\!\!\!\!\!}

\newcommand{\be}{\begin{equation}}  
\newcommand{\ee}{\end{equation}}   
\newcommand{\ba}{\begin{array}}  
\newcommand{\ea}{\end{array}}

\usepackage{diagbox}

\definecolor{rossoCP3}{cmyk}{0,.88,.77,.40}
\definecolor{blueRef}{rgb}{0.2,0.2,0.6}
\definecolor{blue}{rgb}{0,0.396,0.741}
\hypersetup{
	colorlinks, 
	bookmarksopen, 
	bookmarksnumbered,
	citecolor=blueRef, 		
	linkcolor=rossoCP3,	
	urlcolor=rossoCP3,			
}

\newskip\humongous \humongous=0pt plus 1000pt minus 1000pt

\newif\ifdtup

\def\oldreffmt#1{\rlap{[#1]} \hbox to 2\parindent{}}

\def\figfmt#1{\rlap{Figure {#1}} \hbox to 1in{}}  
  
\def\beq{\begin{equation}}  
\def\eeq{\end{equation}}  
\def\bea{\begin{eqnarray}}  
\def\eea{\end{eqnarray}}  
\def\half{\frac{1}{2}}  
  
\def\bq{\begin{quote}}  
\def\eq{\end{quote}}

\def\half{\frac{1}{2}}       

\relax

\newdimen\tdim  
\tdim=\unitlength  
\def\bar{\overline}

\usepackage{color}

\usepackage{graphicx,graphics}

\begin{document}

\preprint{FERMILAB-CONF-22-699}

\title { Gravitational Contact Interactions ---\\
--- Does the Jordan Frame Exist?\\
\vspace{0.2in}
\normalsize{In Honor of Graham G. Ross}\\
}


\author{Christopher T. Hill}
\email{hill@fnal.gov}
\affiliation{Fermi National Accelerator Laboratory\\
P.O. Box 500, Batavia, Illinois 60510, USA\\$ $}

\date{\today}

\begin{abstract}
Scalar--tensor theories, with Einstein Hilbert and  non-minimal interactions, 
$\sim M^2R/2 -\alpha\phi^2R/12 $,
have graviton exchange induced contact interactions.
These always pull the theory back into a formal
Einstein frame in which $\alpha$ does not exist.
In the Einstein frame, 
 contact terms are induced, that are then
absorbed back into other couplings to define the effective
potential (or equivalently, the renormalization group). 
 We show how these effects manifest themselves in a simple model by computing
the effective potential in both frames displaying the discrepancy.
\end{abstract}

\maketitle

\maketitle
\section{Introduction}

Today we are here to honor the life and career of our colleague, Graham Ross.\footnote{
{Invited Talk, "Workshop on the Standard Model and Beyond,\\
Graham G. Ross Memorial Session,''
Corfu, Sept. 1, 2022. }} 
I've known Graham for nearly 50 years, beginning at Caltech when I was a graduate student 
in the mid-1970's and he was a Research Scientist.  
He became a mentor and good friend.

Back then, we worked on the new theory, QCD, exploring the 
photon structure functions, \cite {photon}, and nonleptonic weak interactions, \cite{nonleptonic},
with renormalization group (RG)  methods. 
In the latter 
we followed the pioneering work of \cite{Shifman}, where
I was first introduced to ``contact interactions'' via the so-called ``penguin diagrams'' (so named in
ref.\cite{penguins}).  Contact interactions will play a central role in my talk today.

In 1981, I visited Oxford where Graham had joined the faculty. There we discussed the possibility that
the top quark would be very heavy. He had some ideas based upon the
RG equation for the top quark Yukawa coupling, $g_{top}$. He proposed
that $g_{top}/g_{QCD}$ would flow into the infra-red to a fixed ratio
(a tracker solution), which predicted a top quark mass of about 110 GeV (with Brian Pendleton \cite{PR}).
Note that at this time the experimental limit on the top quark mass was of order $\sim 10 $ GeV, 
and most people believed it would appear at about $\sim 3\times m_{b}\sim 15$ GeV. 

Inspired by the Pendleton-Ross idea, I studied the RG equations 
and found the predicted top quark mass would most probably
be near an ``infrared quasi-fixed point,'' or ``focus point,'' $\sim 220$ GeV \cite{fixed}
assuming standard model running to the high scales. 
People were quite skeptical, however, about a heavy top quark (some people promised 
to ``eat their hats'' if the top mass exceeded $\sim 40 - 90$ GeV
but I don't recall beholding any such feasts). 
In the late 1980's Graham and I overlapped at CERN for a year
and we worked on various generalizations of axions \cite{schiz},
which, particularly in the neutrino sector, proved to have potentially interesting cosmological implications
\cite{late}.

In the 1990's it was becoming clear that the top quark was
surprisingly heavy.  Bardeen, Lindner and I considered
a composite  Higgs boson as a boundstate of top and anti-top quarks \cite{BHL}.
We found that the composite solution implied that $g_{top}$ should be near
the RG quasi--fixed point value, 
inspired from the work of Pendleton and Ross a decade earlier.
The top quark finally weighed in at the Tevatron in 1995 at $175$ GeV,
shy of fixed point prediction by about 20\%.  
I remain of the opinion that this is suggesting an 
intimate dynamical relationship between the top quark and Higgs boson, 
though we may not yet have the full theory (see \cite{CTH3} for a sketch of some new ideas).

Graham Ross was a universal physicist, able to move from phenomenology to theory comfortably.
Above all, he paid close attention to experimental results.   He was a key player in developing
many ideas surrounding the MSSM in the `80's and `90's.   However, circa 2014 he became increasingly interested in alternatives, such
as the Weyl invariant gravity theories which had become an active arena of research \cite{Weyl,authors,HillRoss,staro}.  
We then renewed our collaboration, and that carried us to
his untimely passing.

I feel a great loss,  of a dear friend and colleague,
an amiable person of the highest integrity, an energetic researcher 
with a deep passion for real world physics and its essential
interface with experiment, and
someone I've known for much of my life. 
We leave a half dozen unfinished
papers on our laptops.  
\newpage

 I would like to talk today about one of my more recent works with Graham Ross,
because I feel it is foundational to quantum field theory and model independent. This work
came out of the Covid pandemic, and we had daily multiple email exhanges, 
sometime skype calls, during this time.

The subject of my talk is the field theoretic structure of non-minimal gravitationally
coupled scalars in the regime in which the Planck mass is present (as opposed
to any putative pre-Planckian era).  The question that emerges from
this is: ``Does the Jordan frame really exist?''  You'll see what I
think as the story unfolds.  All of this is described in the paper
 \cite{HillRoss}, and some of the implications are tackled in \cite{staro}.
 A follow-on paper with one of Graham's
 former students, Dumitru Ghilencea, is nearing completion \cite{Ghil}.

\section{Non-Minimal Couplings of Scalars to Curvature}

Over the years there has been considerable interest in Brans-Dicke,
scalar-tensor, and scale or Weyl invariant theories.   
These  have in common
fundamental scalar fields, $\phi_i$,
that couple to
gravity through non-minimal interactions, $\sim M^2R +F(\phi_i)R$.  The Planck mass, $M$
can co-exist with these non-minimal couplings, or the scalars may acquire
VEV's that dynamically generate it \cite{authors}. 
When the theory is proscribed with non--minimal interactions we say it
is given in a ``Jordan frame.''

A key tool in the analysis of these models is the Weyl transformation, \cite{Weyl}.  This involves
a redefinition of the metric, $ g' =\Omega(\phi_i)g$, in which $g$ comingles 
with the scalars. $\Omega$ can be chosen to 
lead to a new effective theory, typically one that has a  pure Einstein-Hilbert action,
  $\sim M^2 R $, in which the non-mimimal interactions 
  have been removed. This is called the ``Einstein frame.'' 
  Alternatively, one might use a Weyl transformation to partially 
  remove a subset of scalars from the non-minimal interactions $\sim M^2 R + F'(\phi_i)R$, 
  where $F'$ is optimized for some particular 
application.

It is {\em a priori}
unclear, however, how or whether the original Jordan frame  theory can be physically equivalent to the
Einstein frame form and how the Weyl transformation is compatible with a full quantum theory 
\cite{Duff}\cite{Ruf}.
Nonetheless, many authors consider this to be a valid transformation
and a symmetry of Weyl invariant theories, and many loop calculations permeate the literature
which attempt to exploit apparent simplifications offered by the Jordan frame.

Graham Ross and I showed that any theory with non-minimal couplings  contains
{\em contact terms} \cite{HillRoss}. These  are generated by the graviton exchange
amplitudes in tree approximation and they are therefore ${\cal{O}}( \hbar^0) $ and 
therefore classical. 
The contact terms occur because emission vertices from the non-minimal interaction are 
proportional to $q^2$ of the graviton, while the Feynman propagator
is proportional to $1/q^2$.  The cancellation of $q^2\times 1/q^2$ 
therefore leads to point-like interactions
that {\em must be included into
the effective action of the theory} at any given order of perturbation theory. 
 The result is that the non-minimal
interactions  disappear from the theory and Planck suppressed higher dimension operators 
appear with modified couplings. 

The structural form of the theory when the contact
term interactions are included corresponds formally to
a Weyl transformation of the metric that takes the theory to the pure Einstein frame
with higher dimension, Planck suppressed, interactions.
In the pure Einstein-Hilbert action there are no classical contact terms,
but at loop level they can be generated, and must be removed as part of the
renormalization group.
However, by virtue of the contact terms, nowhere is a metric redefinition performed
(hence the issue of a Jacobian in the measure of the gravitational
path integral in going to the Einstein frame becomes moot).

 This means that,  provided we are interested in the  theory 
 on mass scales below $M$, 
 the Jordan frame is an illusion and doesn't
 really exist physically.  In the Jordan frame the contact terms are hidden, but they are
 always present, ergo, even though the action superficially appears to be Jordan, 
 it isn't, and remains always in an Einstein frame. 
 
 Efforts to compute quantities, such as effective potentials (or equivalently,
 $\beta$-functions), in the Jordan frame, while ignoring the contact terms,
 will yield incorrect results. Nonetheless, though the non-minimal interactions
 are not present in the classical Einstein frame, they are regenerated
 by loops and the potentials and RG equations are modified by this effect.
 
 In a simple theory in the Jordan frame, where the nonminimal 
 coupling is $-\alpha \phi^2 R/12$,
 this raises the question of how to understand 
 the fate of $\alpha$?  With
 a single scalar with quartic and other interactions, $\lambda_i$, the usual ``naive'' calculation
 of a $\beta$-function in the Jordan frame (``naive'' means ignoring contact terms) yields the form:
 \bea
 \label{one1}
 \frac{\partial \alpha(\mu)}{\partial \ln(\mu)}= \beta_\alpha(\lambda_i) 
 \equiv (1-\alpha)\gamma_\alpha(\lambda_i)
 \eea
 where the factor $(1-\alpha)$ reflects the fact that when $M^2=0$ and  $\alpha=1$ (conformal limit)
 the kinetic term
 of $\phi$ disappers  and $\phi $ becomes static parameter).
 
 However, the contact terms (or a Weyl tranformation) remove $\alpha$  
 and leave an Einstein frame with only the Einstein-Hilbert term, $M^2R$ and the $\phi$ couplings $\lambda_i'$
 (in what follows {\em primed couplings refer the Einstein frame and un-primed to Jordan frame}).
 This means that  $N$ couplings, $(\alpha, \lambda_i)$, in the Jordan 
 frame have become  $N-1$
 couplings, $(\lambda_i')$, in the Einstein frame.  Therefore any physical 
 meaning ascribed to $\alpha$ or
 the $\beta_\alpha$ function is apparantly lost.
 
 Three Feynman diagrams (shown below as D1, D2, D3, in  Figs.(2,3,4)) 
 contribute to $\beta_\alpha$
 in the Jordan frame.  One of them multiplies $\alpha$ in the Jordan frame (D3) and 
 yields the $(1-\alpha)$ factor in eq.(\ref{one1}). 
 But even with $\alpha=0$ in the Einstein frame, two diagrams (D1 and D2) exist and
 reintroduce a perturbative $\delta\alpha$, for a small step in scale $\delta\mu/\mu$.
 This is then removed by the contact terms, but leads to  correction terms in the 
 renormalization of the $\lambda_i'$. 
 We are
 therefore sensitive to the same scale breaking information
 in the Einstein frame that one has in the Jordan frame, 
 which is encoded into $\gamma_\alpha$. 
 This does not, however, imply that the resulting 
 calculations in the Jordan and Einstein frames are then consistent!
 I will explicitly demonstrate the inconsistency through calculation pf
 effective potentials (the RG equations
 of the couplings can always be read off from the effective potentials).
 
 If we stayed in the Jordan frame, with nonzero $\alpha$, and naively
 computed the same effective potential (``naively'' means ignoring contact terms),
 into an Einstein frame, we would obtain a different result.
  The difference 
 is a term proportional to $\alpha$ in the Jordan frame. 
 Equivalently,  going initially to the  Einstein frame 
 and running with the RG, {\em does not commute}
 with running initially in the Jordan frame and subsequently going to the Einstein frame!
 
 We turn presently to a brief discussion of contact terms in general
 and review a simple toy model from \cite{HillRoss} that 
 is structurally similar to the gravitational case.
 We then summarize gravitational contact terms (and refer the reader to \cite{HillRoss}
 for details). We then exemplify the Einstein frame
 renormalization group compared to the naive Jordan frame result, which ignores contact terms,
 and illustrate the discrepancy.

\section{Contact Interactions}

Generally speaking ``contact interactions'' are point-like operators that are
generated in the effective action of the theory in perturbation theory.
They may arise in the UV from ultra-heavy fields that are integrated out, such as the Fermi
weak interaction that arises from integrating out the heavy $W$-boson. 
They may also arise in the IR when a vertex in the theory is proportional to $q^2$ and
cancels against a $1/q^2$ propagator. The gravitational contact term we discuss presently is of the IR form.

\subsection{Contact Terms in Non-gravitational Physics}

Contact terms arise in a number of
phenomena.   Diagrammatically they can arise in the IR when a
vertex for the emission of, e.g., a massless  quantum, of momentum $q_\mu$,
is proportional to $q^2$.  This vertex then cancels the $1/q^2$
from a massless propagator when the quantum is exchanged. This $q^2/q^2$
cancellation leads to an
effective pointlike operator from an otherwise single-particle reducible diagram.

For example, in electroweak physics a
vertex correction by a $W$-boson to a massless gluon emission 
induces a quark flavor changing operator, e.g.,
describing $s\rightarrow d$+gluon, where $s(d)$ is a strange (down) quark.
This has the form of a local operator:
\bea
\label{one}
g\kappa \bar{s}\gamma_\mu T^A d_L D_\nu G^{A\mu\nu}
\eea
where $G^{A\mu\nu}$
is the color octet
gluon field strength and $\kappa \propto
G_{Fermi}$.

This implies a vertex 
for an emitted gluon of 4-momentum $q$ and polarization and color, $\epsilon^{A\mu}$,
of the form
$g\kappa\bar{s}\gamma_\mu T^A d_L  \epsilon^{A\mu}\times q^2 +...$.  However,
the gluon propagates, $\sim 1/q^2$, and couples to a quark current 
$\sim g\epsilon^{A\mu}\bar{q}\gamma_\mu T^A q$.
This results in a contact term:
\bea
\label{localop}&&
g^2 \kappa\left(\frac{q^2}{q^2}\right) \bar{s}\gamma^\mu T^A d_L \bar{q}\gamma_\mu T^A q
\;=\;
g^2 \kappa\bar{s}\gamma^\mu T^A d_L\bar{q}\gamma_\mu T^A q
\nonumbo
\eea
The result is a 4-body local operator 
which mediates electroweak transitions
between, e.g., kaons and pions \cite{Shifman}, also
known as ``penguin diagrams'' \cite{penguins}.
Note the we can rigorously obtain the contact term result
by use of the gluon field equation within the operator
of eq.(\ref{one}), 
\bea
D_\nu G^{A\mu\nu}= g\bar{q}\gamma^\mu T^A q.
\eea
This is justified as operators that vanish by equations of motion, 
known as ``null operators,'' will generally have
gauge noninvariant anomalous dimensions and are unphysical 
\cite{Deans}.

Another example occurs in the case of
a cosmic axion, described by an oscillating  classical 
field,  $\theta(t)=\theta_0\cos(m_at)$, 
interacting with a magnetic moment, $ \vec{\mu}(x)\cdot \vec{B}$, through the
electromagnetic anomaly $ \kappa \theta(t) \vec{E}\cdot \vec{B} $. A static magnetic moment 
emits a virtual spacelike photon of momentum $(0,\vec{q})$. The anomaly 
absorbs the  virtual photon and emits an on-shell photon of polarization $\vec{\epsilon}$,
inheriting energy $\sim m_a$ from the cosmic axion.
The Feynman diagram, with the exchanged virtual
photon, yields an amplitude, 
$\propto (\theta_0\mu^i\epsilon_{ijk}q^j)(1/\vec{q}^{\;2}) (\kappa \epsilon^{k\ell h}q_\ell m_a\epsilon_h)
\sim  (\kappa\theta_0 m_a\vec{q}^{\;2}/\vec{q}^{\;2}) \vec{\mu}\cdot \vec{\epsilon}$. 
The $ \vec{q}^{\;2}$ factor then 
cancels the $1/\vec{q}^{\;2}$ in the photon propagator,
resulting in a contact term which is an induced, parity violating,
oscillating electric dipole interaction:
$\sim \kappa \theta (t) \vec{\mu}\cdot \vec{E}
$. This results in 
cosmic axion induced {\em electric dipole} radiation from
any magnet, including an electron \cite{CTHa}.

\subsection{Illustrative Toy Model of Contact Terms}

To illustrate the general IR contact term phenomenon,
consider a single massless real scalar field $\phi $
and operators $A$ and $B,$ which can be functions of other fields, with the 
action given by:
\bea
S=\int \frac{1}{2}\partial \phi \partial \phi -A\partial ^{2}\phi -B\phi 
\label{toy0}
\eea
Here
$\phi $ has a propagator ${i}/{q^{2}}$, but the vertex of a diagram
involving
$A$ has a factor of $\partial^2 \sim -q^{2}$. This yields a pointlike interaction, $\sim
q^{2}\times ({i}/q^{2})$,
in a single particle exchange of $\phi $, and therefore implies contact terms.

At lowest order in perturbation theory consider the diagram with $\phi$
exchange in Fig.(1).  This involves two time-ordered products
of interaction operators: 
\bea
&& 
 {T} \;\;  i\!\!\int \! A\partial ^{2}\phi \times i\!\!\int \! B\phi 
 \rightarrow \frac{iq^{2}}{q^{2}}\
 AB\;\;\;=\; i\int \! AB
\nonumbo 
\frac{1}{2}{T} \;\;  i\!\!\int \! A\partial ^{2}\phi \times
i\!\!\int \! A\partial ^{2}\phi 
 \rightarrow 
\frac{i(q^{2})^{2}}{2q^{2}}A^2   
\!=\!\frac{i}{2}
\int \! A\partial ^{2}A.
\nonumbo
\eea
where $d^4x$ is understood in the integrals
amd the $\half$ factor in the  $A\partial^2A$ term comes from the second order
in the expansion of the path integral $\exp (i\int A\partial^2 \phi)$. Note that
we also produce a nonlocal  interaction $ -{i}B^2/{2q^{2}}.$ 

\begin{figure}[t!]
\vspace{0.0 in}
	\hspace*{0.1in}\includegraphics[width=0.75\textwidth]{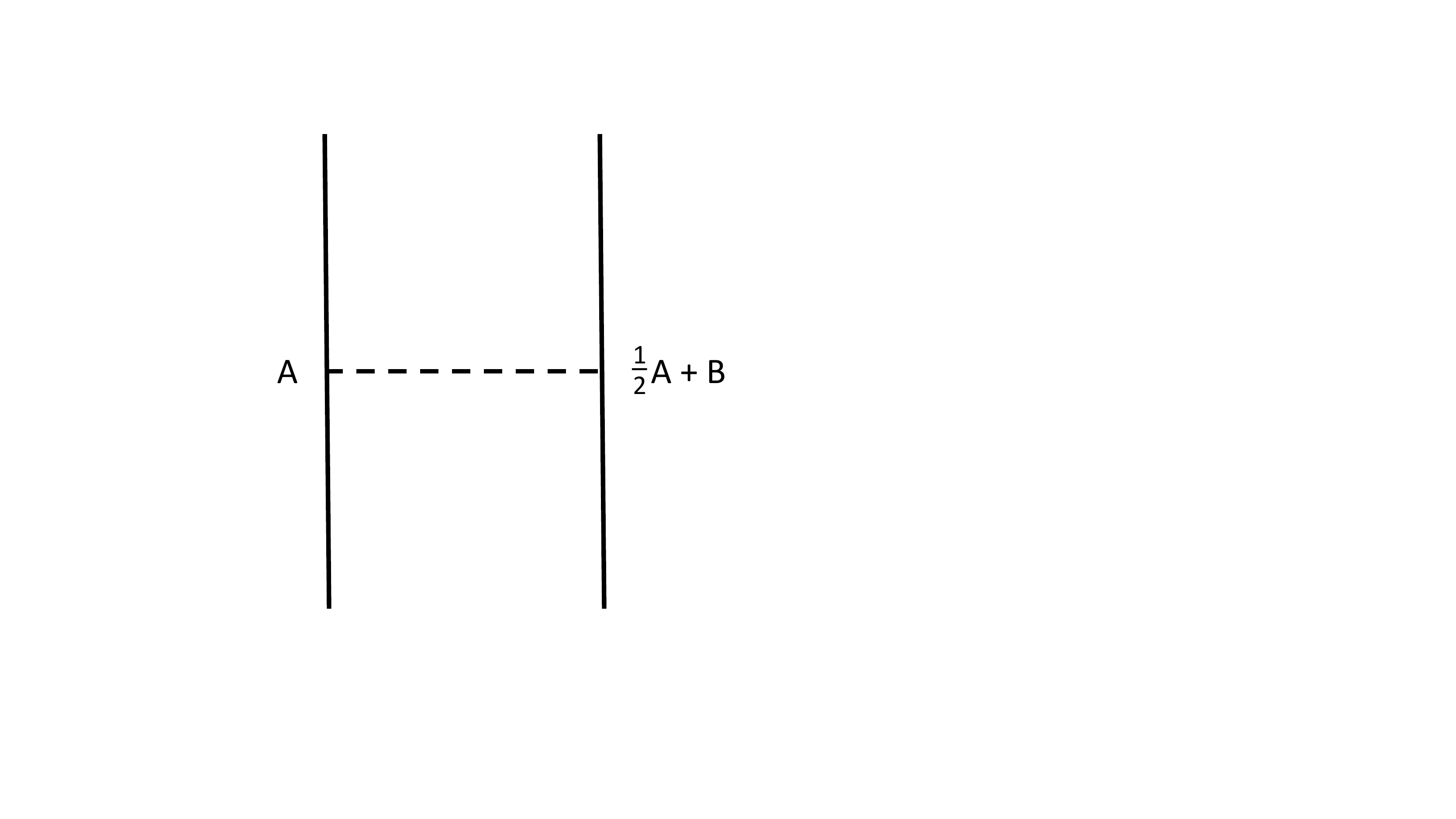}
	\vspace{-0.5 in}
	\caption{Contact terms in the toy model are
	generated by diagrams with exchange of $\phi$ (dashed). In gravity,
	with non-minimal term $\sim \int\! \sqrt{-g} F\left(\phi\right) R$ 
	and matter field Lagrangian $\sim \int\! \sqrt{-g} L\left(\phi\right)$
	then $A$ is replaced by $F(\phi)$ and $B$ is replaced by $L(\phi)$, and the dashed line
	is a graviton propagator. }
\end{figure}
We thus see that we have
diagrammatically obtained a local effective action:
\bea
S=\int \frac{1}{2}\partial \phi \partial \phi +\frac{1}{2}A\partial
^{2}A+AB +\makebox{ {long distance}}
\eea
Of course, we can see this straightforwardly by
``solving the theory,'' by defining a shifted field:
\bea
\phi =\phi^{\prime } -\frac{1}{\partial ^{2}}\left( \partial ^{2}A+B\right) 
\eea
Substituting this into the action $S$  and integrating by parts yields:
\bea
\label{eff1}
S =\int \frac{1}{2}\partial \phi ^{\prime }\partial \phi ^{\prime }+\frac{%
1}{2}A\partial ^{2}A+AB+\frac{1}{2}B\frac{1}{\partial ^{2}}B
\eea
An equivalent effective local action that describes both
short and large distance is then,
\bea
\label{efftoy}
S=\int \frac{1}{2}\partial \phi \partial \phi +\frac{1}{2}%
A\partial ^{2}A+AB-{B \phi }
\eea
The contact terms have become pointlike components of the effective action,
while the remaining
{long} distance effects are produced by the usual massless $\phi $ exchange.  Note that the
derivatively coupled
operator $A$ has no long distance interactions {due to} $\phi $ exchange. 
Moreover, in the
effective action of
eq.(\ref{efftoy}) we have implicitly ``integrated out'' the $A\partial ^{2}\phi $, which is no
longer part of the action and is replaced
by new operators $\frac{1}{2}A\partial ^{2}A+AB$.
We will see that this is exactly what happens with gravity, where
the $A\partial^{2}\phi$ term is schematically the nonminimal $F(\phi)R(g)\sim F(\phi) \partial^2 h$
term in a weak field expansion of gravity $g =\eta+h$.

One can also adapt the use of equations of motion to obtain eq.(\ref{efftoy})
from the action eq.(\ref{toy0}) but
this requires care. For example, the insertion of the $\phi$ equation of motion,
into $A\partial^2\phi$ correctly gives the $AB$ term but misses the 
factor of $1/2$ in the $A\partial^2A$ term.  We can therefore do a trick of
defining a
``modified equation of motion,'' where we supply a factor of $1/2$ on the term  $A\partial^2A$,
e.g. substitute:
\bea
\label{subst}
\partial ^{2}\phi=-\partial ^{2}A-B \rightarrow  -\frac{1}{2}\partial ^{2}A-B
\eea
in place of the $ \partial ^{2}\phi$ in the second term of eq.((\ref{toy0}))
to obtain eq.(\ref{efftoy}).

\section{Gravitational Contact Terms}

Consider a general theory involving scalar fields $\phi _{i},$
an Einstein-Hilbert {term and} a non-minimal interaction:
\bea
\label{action00}
&&\backo\;\;
S=\int \sqrt{-g}\left(\frac{1}{2}M^{2}R\left( g_{\mu \nu }\right) +\frac{1}{2%
}F\left( \phi _{i}\right) R\left( g_{\mu \nu }\right) +L\left( \phi
_{i}\right) \right)
\nonumbo
= S_1 + S_2 +S_3
\eea
where we use the 
metric signature and curvature tensor conventions of \cite{CCJ}.
 $S_{1}$ is the kinetic term of gravitons
\bea
S_{1} &=& \frac{1}{2}M^{2}\int \sqrt{-g}R
\eea
and becomes the Fierz-Pauli action in a weak field expansion.

$S_{2}$ is the non-minimal interaction, and takes the form:
\bea
S_{2}=\frac{1}{2}\int \sqrt{-g}F\left( \phi _{i}\right) R\left( g_{\mu \nu }\right)
\eea

$S_{3}$ is the matter action with couplings to the gravitational weak
field:
\bea
S_{3}=\int \sqrt{-g}\;L\left( \phi _{i}\right) 
\eea
The Lagrangian takes the form
\bea
L\left( \phi_{i}\right)  = \frac{1}{2}g^{\mu \nu }\partial _{\mu }\phi_i \partial _{\nu }\phi_i-W(\phi_i)
\eea
with potential $ W(\phi_i) $.
The matter lagrangian has stress tensor and stress tensor trace:
\bea
T_{\mu \nu }&=&\partial _{\mu }\phi_i \partial _{\nu }\phi_i -g_{\mu \nu }\left( 
\frac{1}{2}g^{\rho \sigma }\partial _{\rho }\phi_i \partial _{\sigma }\phi_i
-W(\phi_i )\right) .
\nonumber \\
T&=&-\partial ^{\sigma }\phi_i \partial _{\sigma
}\phi_i +4W(\phi_i)
\eea
There are then three ways to obtain the contact term:

 \vspace{0.2in}
\noindent
 (1) GRAVITON EXCHANGE CONTACT TERM
 
 \vspace{0.2in}
 \noindent

In reference \cite{HillRoss} 
the corresponding Feynman diagrams of Figure 1 with are evaluated arising from single graviton exchange
between the interaction terms.  

We treat the theory perturbatively, expanding around flat space. {Hence} we
linearize gravity with a weak field $h_{\mu \upsilon }$:
{\bea
\label{linear}
g_{\mu \upsilon }\approx \eta _{\mu \upsilon }+\frac{h_{\mu \upsilon }}{%
M}.
\eea}
The scalar curvature is then:
\bea
R &=& R_{1}+R_{2}
\nonumber \\
M R_{1}&=&\biggl( \partial^{2}h\ -\partial^{\mu}\partial ^{\nu}h_{\mu \nu }\biggr) 
\nonumber \\
M^{2}R_{2}&=&  -\frac{3}{4}\partial ^{\rho }h^{\mu
\nu }\partial_{\rho}h_{\mu \nu }
-\frac{1}{2}h^{\mu
\nu }\partial ^{2}h_{\mu \nu }+... 
\eea
(see \cite{HillRoss} for the complete expression for $R_2$).
 $S_{1}$ then becomes:
\bea &&
\label{FP0}
S_{1} = \frac{1}{2}M^{2}\int \sqrt{-g}R
= \frac{1}{2}%
M^{2}\int\!\! \left( R_1+R_{2}+\frac{1}{2}\frac{h}{M}R_{1}\right) 
\nonumbo
= \frac{1}{2}\int h^{\mu \nu }\biggl( \frac{1}{4}\partial ^{2}\eta _{\mu \nu
}\eta _{\rho \sigma }-\frac{1}{4}\partial ^{2}\eta _{\mu \rho }\eta _{\nu
\sigma }
\nonumbo \qquad\qquad
-\frac{1}{2}\partial _{\rho }\partial _{\sigma }\eta _{\mu \nu }+%
\frac{1}{2}\partial _{\mu }\partial _{\rho }\eta _{\nu \sigma }\biggr)
h^{\rho \sigma }.
\eea
 Note that the leading term, $\int R_{1}$,
is a total divergence and is therefore zero in the Einstein-Hilbert
action, and what remains of eq.(\ref{FP0})
is the Fierz-Pauli action.   This is key to the origin of the contact terms.
The non-minimal interaction, $S_{2}$, then takes the leading form:
\bea &&
S_{2}=\frac{1}{2}\int \!\!\sqrt{-g}\; F\left( \phi\right) R\left( g\right)
\rightarrow \frac{1}{2}\int \!\!F\left( \phi\right) R_1 \left( g\right)
\nonumbo
\qquad
=
\int\frac{1}{2M}F\left( \phi\right) 
\Pi
^{\mu \nu }h_{\mu \nu } 
\eea
where it is useful to introduce the transverse derivative,
\bea
\Pi ^{\mu \nu }= \partial ^{2}\eta ^{\mu \nu }\ -\partial
^{\mu }\partial ^{\nu } .
\eea 
$S_{2}$ involves derivatives, 
and is the analogue of the $A\partial^2\phi$ term in eq.(\ref{toy0}).
It will therefore generate
contact terms in the gravitational potential due to single graviton
exchange. $S_3$ is the analogue of the $B\phi$ term in eq.(\ref{toy0}), and 
this situation will 
closely parallel the toy model.

In \cite{HillRoss} we developed the graviton propagator, following
the nice lecture notes of
Donoghue {\it et. al.,} \cite{Donoghue}.
We remark that we found a particularly
useful gauge choice,
\bea
\label{wgauge}
\partial_\mu h^{\mu \nu} = w \partial^\nu h
\eea
where $w$ defines a single parameter family
of gauges.
The familiar De Donder gauge corresponds to $w=\frac{1}{2}$, 
while the  choice $w=\frac{1}{4}$ is particularly natural in
this application, and the gauge invariance
of the result is verified by the $w$-independence
(we verify the Newtonian potential from graviton
exchange between static masses in $w$ gauge; see \cite{HillRoss}).

We can the compute single graviton exchange between the interaction
terms of the theory.
A diagram with a single
$S_2$ vertex and single $S_3$ vertex is the analogue of $AB$ in the toy model and yields:
\bea
-i\left\langle  T\;S_{2}S_{3}\right\rangle=
\int d^{4}x\; \frac{F\left( \phi _{i} \right) }{2M^{2}}%
T( \phi _{i}) 
\eea
Also we have 
the pair $\left\langle S_{2}S_{2}\right\rangle $
which corresponds to $\frac{1}{2} A\partial^2 A$ in the toy model and yields:
\bea 
-i{\left\langle T\; S_{2}S_{2}\right\rangle}&=&-\int d^{4}x\;  \frac{3}{4M^{2}}\; F\left(
\phi _{i} \right) \!\partial ^{2}F\left( \phi _{i}\right) 
\nonumbo
\eea
Hence,  the action becomes
\bea
 &&
S = S_1+S_3+ S_{CT}
\eea
where
\bea
\label{CT}
&&
S_{CT}= 
\int d^{4}x \biggl( 
-\frac{3}{4M^{2}} F \partial ^{2}F 
+\frac{1 }{2M^{2}}FT
\biggr)
\eea
Note the sign of the $F\partial ^{2}F$ is opposite (repulsive) to that of
the toy model $A\partial^2 A$.  Since this is a tree diagram effect,
it is classical.

\vspace{0.2in}
\noindent
(2) WEYL TRANSFORMATION 
\vspace{0.2in}
 \noindent

In eq.(\ref{action00}) we define:
\bea
\Omega ^{2}=\left( 1+\frac{F\left( \phi_i \right) }{M^{2}}\right)
\eea
and 
perform a Weyl transformation on the metric:
\bea
\label{metricweyl}
&&
g_{\mu \nu }(x)\rightarrow \Omega ^{-2}g_{\mu \nu }(x)
\qquad
g^{\mu\nu }(x)\rightarrow \Omega ^{2}g^{\mu \nu }(x)
\nonumbo
\sqrt{-g}\rightarrow\sqrt{-g}\Omega^{-4}
\nonumbo
R(g)\rightarrow\Omega ^{2}R(g')+6\Omega ^{3}D\partial \Omega ^{-1}
\eea
and the action of eq.(\ref{simp}) becomes:
\bea &&
S\rightarrow\int \sqrt{-g}\biggl(\frac{1}{2}M^{2}
R\left( g\right) 
\nonumbo
-3M^{2}\partial_\mu \left( 1+\frac{F}{
M^{2}}\right)^{1/2}\!\!\partial^\mu 
\left( 1+\frac{F}{M^{2}}\right)
^{-1/2} 
\nonumbo
 +\frac{1}{2}\left(1+\frac{F}{M^{2}}\right)^{-1}\!\!\!
\partial _{\mu }\phi_i \partial^{\nu }\phi^i 
-\left( 1+\frac{F}{M^{2}}\right)^{-2}\!\!\!W(\phi_i)\biggr)
\nonumbo
\eea
Keeping terms to $O({1\over M^2})$ and integrating by parts we have:
\bea && \backo
S=S_{1}+S_3
\nonumbo \backo
+\int d^4x\bigg(-\frac{3
F\left( \phi _{i}\right)\partial ^{2}F\left( \phi _{i}\right)}{4M^{2}}
+\frac{F\left( \phi _{i}\right)T\left( \phi _{i}\right)}{2M^{2}}\bigg) 
\eea
The Weyl transformed action is identically consistent with the contact terms
of eq.(\ref{CT}) above,
to first order in $1/M^2$.

Hence, contact terms arise in gravity with non-minimal
couplings to scalar fields due to graviton
exchange and their form is equivalent to
a Weyl redefinition of the theory to the  Einstein frame.
However we emphasie that {\em the contact terms are not a Weyl tranformation}
since they do not involve a rescaling of the metric, and there 
would therefore be no Jacobian ghost terms introduced into the path integral.
 Hence any theory
with a non-minimal interaction $\sim F(\phi)R$ will lead
to  contact terms at order $1/M^2$.  What we say presently only applies
on scales below $M$, hence the Jordan frame
would, at best, apply only in a UV completion of the theory to pre-Planckian scales.

 \vspace{0.2in}
\noindent
(3)  USE OF $R$ (MODIFIED) EQUATION OF MOTION
\vspace{0.2in}
 \noindent

The Einstein equation with the nonminimal term is:
\bea &&
M^{2} 
G_{\alpha \beta }=-T_{\alpha \beta }-D_\mu(D_\nu F(\phi_i)) + g_{\mu\nu} D^2 F(\phi_i)
\nonumbo
M^2 R=  T - 3 D^2 F(\phi_i)
\eea
We can use a simple trick to obtain the contact term via
a modified the equation of motion for $R$.  We  supply a factor of $1/2$
in the last term which is the analogue of the $\partial^2 A$ term
as in eq.(\ref{subst}):
\bea 
\label{RE0} &&
 R'=\frac{1}{M^2} \biggl( T - \frac{3}{2} D^2 F(\phi_i) \biggr)
\eea
Then
substitute $R'$ for $R$  in the non-minimal term $F R$ of the action of eq.(\ref{simp})
\bea &&\backo \;\;
S\rightarrow\int \sqrt{-g}\left(\frac{1}{2}M^{2}R\left( g_{\mu \nu }\right) +\frac{1}{2%
}F\left( \phi _{i}\right) R'\left( g_{\mu \nu }\right) +L\left( \phi
_{i}\right) \right)
\nonumbo
 \qquad =S_{1}+S_3
\nonumbo
\qquad
+
\int d^4x\bigg(-\frac{3
F\left( \phi _{i}\right)\partial ^{2}F\left( \phi _{i}\right)}{4M^{2}}
+\frac{F\left( \phi _{i}\right)T\left( \phi _{i}\right)}{2M^{2}}\bigg)
\nonumbo
\eea
In the RG calculation we will only need the exact equation of motion 
for $R$ in the Einstein frame (without the pseudo $-3D^2F/2$ term), 
so this ambiguity does not arise.

\section{ Effective Potetial
in a Simple Model}

Consider the following action:
\bea
\label{simp}
&&\backo
S_{Jordan}=\int \sqrt{-g}\biggl( \frac{1}{2}\partial\phi\partial\phi
-\frac{\lambda _{1}}{4}\phi ^{4}-
\frac{\lambda _{2}\phi ^{6}}{12M^{2}}
\nonumbo
-\frac{\gamma }{12}\frac{\phi ^{2}}{M^{2}}\partial\phi\partial\phi
-\frac{\alpha}{12}\phi ^{2}R
+\frac{1}{2}M^{2}R
\biggr) 
\eea
This is the most general action for a real scalar field with a $Z_2$ symmetry $\phi\rightarrow -\phi$
valid to O($M^{-2}$) with Einstein gravity and assuming $\phi$ is massless, $m^2=0$.
Here we do not include a term $\phi^4R/M^{2}$ since, after use of equations of motion, $R\sim M^{-2}$
such a term would enter  the physics at O($M^{-4}$).

We can go to the Einstein frame by implementing the CT.
We find that the
effect of single graviton exchange to eq.(\ref{simp})
yields a new action to order $M^{-2}$:
\bea
\label{simp2}
&&\backo
S_{Einstein}=\!\int\! \sqrt{-g}\biggl( \frac{1}{2}\partial\phi\partial\phi
-\frac{\lambda' _{1}}{4}\phi ^{4}
-
\frac{\lambda'_2\phi ^{6}}{12M^{2}}
\nonumbo 
-\frac{\gamma'\phi ^{2} }{12M^{2}}\partial\phi\partial\phi
+\frac{1}{2}M^{2}R 
\biggr)
\eea
where:
\bea \label{stuff0}
&&
\gamma' =\gamma -\alpha-\alpha^{2}
\nonumbo
\lambda_1' = \lambda_1 
\nonumbo
\lambda_2'=\lambda_2+\alpha\lambda_1
\eea
We see that to first order in $M^{-2}$ in $S_{Einstein}$
we have three interaction terms,  though the original
action $S_{Jordan}$ displayed four interaction terms. In the Einstein frame action
we see that $\alpha$ has disappeared having been absorbed into
redefining the other coupling constants. 
This indicates that the nonminimal term in $S_{Jordan}$ with coupling 
$\alpha$ is unphysical. Moreover, 
if we are careful to incorporate the effects of the contact term, 
then  $S_{Einstein}$ will ``close'' under renormalization.

We define,
\bea
Z = 1-\frac{\gamma' \phi^2}{6M^2}
\eea
and the Einstein action becomes:
\bea
&&
\label{}
S_{Einstein}=
\!\int\! \sqrt{-g}\biggl( \frac{1}{2}Z\partial\phi\partial\phi
-\frac{\lambda' _{1}}{4}\phi ^{4}
-
\frac{\lambda'_2\phi ^{6}}{12M^{2}}\nonumbo 
+\frac{1}{2}M^{2}R 
\biggr)
\eea
 We  do a
background field calculation where we 
shift $\phi\rightarrow \phi_0+\sqrt{\hbar}Z_0^{-1/2}\hat{\phi}$,
where
we define
\bea &&
Z_0 = 1-\frac{\gamma' \phi_0^2}{6M^2}
\eea
We will treat $\phi\rightarrow \phi_0$ as a {\em constant} and
compute the one loop, $\cal{O(\hbar)}$ effective potential in $\phi_0$
integrating out the quantum fluctuations $\hat{\phi}$ .\footnote{ In a 
forthcoming paper with Ghilenca \cite{Ghil}
we will give the  general effective action where the backgound field
is treated as non-constant and dynamical.  The present analysis parallels
a Coleman-Weinberg potential \cite{CW}.}
Expand the action with the
shifted field to O$(\hat\phi^2)$, dropping terms odd in $\hat{\phi}$
for constant $\phi_0$.
\bea 
\label{}
&&\backo
S_{Einstein}
=
\nonumbo\backo
\int \sqrt{-g} \biggl( \frac{1}{2} \partial \hat{\phi} \partial\hat{\phi}
-\half B \hat\phi^2 
-V_0(\phi_0)
+\frac{1}{2}M^{2}R
 \biggr) 
 \nonumbo
\eea
where,
\bea &&
\label{rels}
Z_0 = 1-\frac{\gamma' \phi_0^2}{6M^2}
\nonumbo
B=Z_0^{-1}\biggl({3\lambda'_1}\phi_0^{2}
+\frac{5\lambda' _{2}}{2M^2}\phi_0^4\biggr)
\nonumbo
={3\lambda'_1}\phi_0^{2}
+\frac{5\lambda'_{2}+ \gamma'\lambda'_{1}}{2M^2}\phi_0^4
\nonumbo
V_0(\phi_0)=\frac{\lambda' _{1}}{4}\phi_0^{4}
+\frac{\lambda' _{2}}{12}%
\frac{\phi_0 ^{6}}{M^{2}}
\nonumbo
\eea
First we consider the non-curvature terms, with flat Minkowsky space
metric $g_{\mu\nu}=\eta_{\mu\nu}$.
We obtain the effective potential from the log of the path integral
 described briefly in the Appendix, 
 eq.(\ref{19}), obtained by integrating out $\hat{\phi}^2$.
 Presently we plug $m^2\rightarrow B$ into the path integral expression of eq.(\ref{19}),
 \bea
\label{19c}
\Gamma_0
&=&
-\frac{1}{2}  B^{2} L   +O \biggl(  \frac{B^{3}}{\Lambda^{2}} \biggr)    
\eea
and we define the log as:
\bea
L=\frac{1}{16\pi^{2}}\ln\frac{\Lambda}{\mu}
\eea
with a generic infrared cut-off mass scale $\mu$.
Hence, squaring $B$ to  {\cal{O}}($\hbar$), the resulting potential is
to ${\cal{O}}(\phi_0^6)$:
\bea &&
\label{20}\backo \!\!
\Gamma_0 =
-\biggl(
\frac{9\lambda'^2_1}{2} \phi_0^{4}
+\frac{(15\lambda'_2\lambda'_1+3\gamma'\lambda'_1{}^2) }{2}\frac{\phi_0^6}{M^{2}}
\biggr)   L
\eea
Note that  $\mu\rightarrow \phi_0$ in the logarithm
this is essentialy a Coleman-Weinberg potential \cite{CW}.

\subsection{ Inclusion of Gravitational Effects}

Consider the weak field approximation to gravity. 
We choose $w$-gauge of eq.(\ref{wgauge}) and obtain
\bea
\label{grav1}
\partial _{\alpha }h^{\alpha \beta }=w\partial
^{\beta }h, \qquad  R=(1-w)\partial ^{2}h/M
\eea
Up to linear terms in $h_{\mu\nu}\hat{\phi}^2/M$ we have,
\bea &&
S_{Einstein}\rightarrow \int \Biggl( \half \partial_{\mu }\hat\phi\partial^{\mu }\hat\phi
+\frac{1}{4}\frac{h}{M}\partial_{\mu }\hat\phi\partial^{\mu }\hat\phi+\frac{1}{2}M^{2}R
\nonumbo 
-\frac{h^{\mu \upsilon }}{2M}  \partial_{\mu }\hat\phi\partial _{\nu }\hat\phi
-\half \biggl(1+\frac{1}{2}\frac{h}{M}\biggr) B\hat\phi^{2}-V_0(\phi_0)  \Biggr)
\eea
The contributions to the potential from the Feynman diagrams
of Figs.(1,2) are then,
\bea &&
\backo\backo 
\Gamma_{D1}
 =-\frac{ \bigl( 1+2w \bigr) }{12M}B \partial ^{2}h L
\nonumbo
\backo\backo 
\Gamma_{D2}
 =\frac{B}{4M}\partial^{2}h  L
\eea
Full details will be given in \cite{Ghil}.
Noting that,
\bea &&
\frac{1}{4M} \partial^{2}h
-\frac{ ( 1+2w) }{12M} \partial ^{2}h 
=\frac{( 1-w )}{6M}\partial ^{2}h
=\frac{1}{6M}R
\nonumbo
\eea
Hence we have the net contribution to the potential,
\bea
\label{D1D2}
&& \Gamma_\alpha\equiv
\Gamma_{D1}+\Gamma_{D2}
= \frac{1}{6}BR \;L=\frac{\lambda_1\phi_0^2}{2}R L
\eea
We thus see that  there is a $\delta\alpha\phi_0^2 R/12$ term generated  from the
$3\lambda'_1\phi_0^2$ term in B: 
 \bea &&
 \delta\alpha = 6\lambda'_1 L +O(1/M^2)
 %
 \eea
 Note that there are other terms proportional to $R$, but
 we only keep leading terms in $1/M^2$ in $\delta\alpha$, since $R\sim 1/M^2$.
 
To implement the contact term we
use, in eq.(\ref{D1D2}), the leading order $R$ equation of motion in the Einstein frame,
\bea
\label{R00}&&
R=\frac{1}{M^2}  T=\frac{1}{M^2}4\times \biggl(\frac{\lambda_{1}}{4}\phi_0^{4}-\frac{9\lambda^2_{1}}{2}\phi_0^{4}L
 \biggr)+{\cal{O}}\biggl(\frac{1}{M^4}\biggr)
 \nonumbo
 \rightarrow \lambda_1 \frac{\phi_0^{4}}{M^2}+{\cal{O}}(\hbar)
\eea
We then have
 \bea
 && \Gamma_\alpha
=\frac{1}{2}\lambda'_1\phi^2_0  R \;L
 \rightarrow\frac{\lambda'^2_1\phi_0^6}{2M^2}L + 
{\cal{O}}\frac{1}{M^{4}}
 \eea
Therefore, combining all effects
\bea&&
\label{resultE}
\Gamma_{Einstein}\equiv
V_0+\Gamma_0+\Gamma_\alpha =
\nonumbo
\frac{\lambda _{1}}{4}\phi_0^{4}
+\frac{\lambda _{2}}{12}%
\frac{\phi_0 ^{6}}{M^{2}}
\nonumbo
-\half \biggl(
9\lambda^2_1 \phi_0^{4} +(15{\lambda_1\lambda_2 }+3\lambda^2_1\gamma -[\lambda_1^2])\frac{\phi_0^6}{M^{2}}
\biggr) L
\eea
where the term in $[..]$ comes from the gravitational effects of D1 and D2.

\begin{figure}[t!]
\vspace{0 in}
	\hspace*{-0.1in}\includegraphics[width=0.5\textwidth]{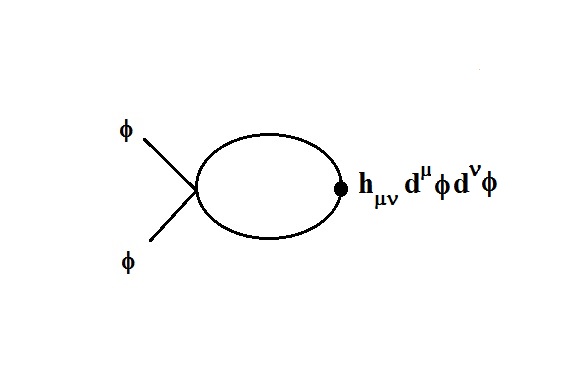}
	\vspace{-0.5 in}
	\caption{ Diagram D1. }
\end{figure}
\begin{figure}[t!]
\vspace{0 in}
	\hspace*{-0.1in}\includegraphics[width=0.5\textwidth]{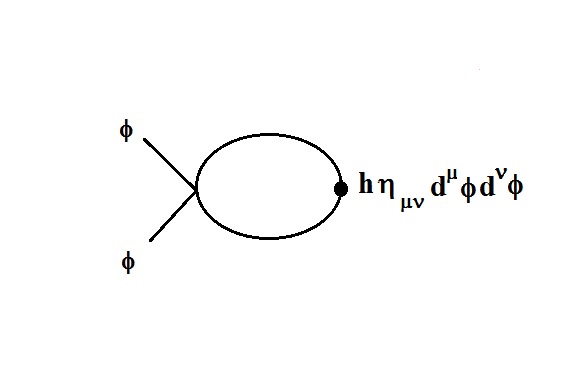}
	\vspace{-0.5 in}
	\caption{ Diagram D2.}
\end{figure}

From eq.(\ref{resultE}) we can extract the $\beta$-functions for $\lambda_1$ and $\lambda_2$.
Define $D= 16\pi^2 \partial/\partial \ln\mu$, and we have:
\bea &&
D\lambda_1 = 18\lambda_1^2
\nonumbo
D\lambda_2=90\lambda_1\lambda_2-6\lambda_1^2+18\lambda^2\gamma'
\eea
In \cite{Ghil} we obtain the RG equation for $\gamma'$, but this
requires an effective action for non-constant $\phi_0$.

\subsection{Comparison to a Conventional Calculation of  $\beta_\alpha$ \\
in Jordan Frame Neglecting Contact Terms}

We now compute the potential in the Jordan frame where
we neglect the contact term, as is conventionally done.  Here we
again use the background field method.

Return to  $S_{Jordan}$, and begin by shifting $\phi\rightarrow \phi_0+\sqrt{\hbar}\hat{\phi}$
and expanding to O$\hat\phi^2$
where $\phi_0$ is a constant classical background field and $\sqrt{\hbar}\hat{\phi}$
is a quantum fluctuation. Henceforth we will suppress the factor of $\sqrt{\hbar}$
but we will compute loops to order $\hbar$.  
The shifted action 
where we drop the linear terms in $\hat{\phi}$ becomes to order $\hbar$:
\bea 
\label{simpR0}
&&
S_{Jordan}
=\int \sqrt{-g} \biggl( \frac{1}{2}g^{\mu \upsilon }\partial _{\mu} \hat{\phi}\partial _{\nu } \hat{\phi}
\nonumbo 
-\half Z_0^{-1}A\hat\phi ^{2}R
-\half B \hat\phi^2 
-V_0(\phi_0)
+\frac{1}{2}M^{2}R
 \biggr) 
\eea
where, we have the relations of eq.(\ref{rels}), but with unprimed couplings replacing the 
primed ones, and
\bea &&
A=\frac{\alpha}{6}
\eea
Expanding to linear terms in $h_{\mu\nu}/M$ 
in weak field gravity 
the action eq.(\ref{simpR0}) becomes
\bea
\label{A4}&&
S\rightarrow \int \Biggl( \half  \partial_{\mu }\hat\phi\partial^{\mu }\hat\phi
+\frac{h}{4M}\eta^{\mu\nu}\partial_{\mu }\hat\phi\partial_{\nu }\hat\phi
-\frac{h^{\mu \upsilon }}{2M}  \partial_{\mu }\hat\phi\partial _{\nu }\hat\phi
\nonumbo 
-\half Z_0^{-1} AR{\hat\phi^{2}}
- \half B\hat\phi^{2}-B\frac{h}{4M}\hat\phi^{2}-V_0(\phi_0)+\frac{1}{2}M^{2}R \Biggr)
\nonumbo
\eea
\begin{figure}[t!]
\vspace{0 in}
	\hspace*{-0.1in}\includegraphics[width=0.5\textwidth]{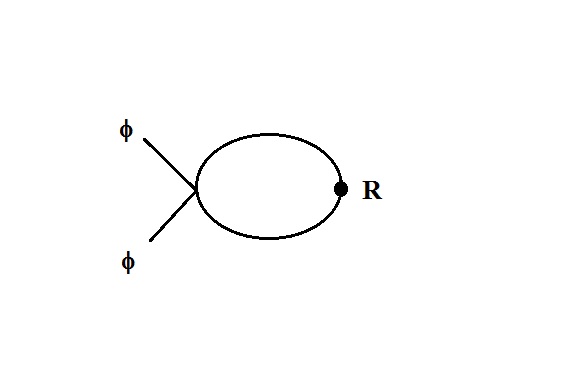}
	\vspace{-0.5 in}
	\caption{ Diagram D2.}
\end{figure}

Neglecting the contact terms we see, in addition to the non-curvature potential
obtained previously in eq.(\ref{20}),
the action eq.(\ref{A4}) now (naively) generates three diagrams linear in the curvature,
 D1, D2, and D3 of Figs(2,3,4). 
The additional D3 diagram (which is absent in the Einsteain frame) is:
\bea &&
\label{D3}
\backo\backo
\Gamma_{D3}
=-Z_0^{-1} ABR L =-\frac{\alpha}{6}BRL +O\frac{1}{M^2}R
\eea
Hence we have from eqs.(\ref{D1D2},\ref{D3}) for (D1+D2+D3):
\bea &&
\Gamma_{D3}+\Gamma_\alpha
= -\frac{1}{6}( \alpha-1 )  BR\;L 
\nonumbo
= 
-\frac{1}{6}( \alpha-1)
\biggl( {3\lambda_1}\phi_0^{2}
+\frac{5\lambda_{2}+ \gamma\lambda_{1}}{2M^2}\phi_0^4\biggr) RL
\eea
Therefore, combining all effects:
\bea &&
\Gamma_{Jordan} =
\nonumbo
\frac{\lambda _{1}}{4}\phi_0^{4}
+\frac{\lambda _{2}}{12}
\frac{\phi_0 ^{6}}{M^{2}}+\frac{\alpha}{12}\phi_0^2R+\frac{\gamma}{12M^2}\phi_0^2R
\nonumbo
-\half L \biggl(
9\lambda^2_1 \phi_0^{4} +(15{\lambda_1\lambda_2 }+3\lambda^2_1\gamma)\frac{\phi_0^6}{M^{2}}
\biggr) 
\nonumbo
-\frac{1}{6}( \alpha-1)
\biggl( {3\lambda_1}\phi_0^{2}
+\frac{5\lambda_{2}+ \gamma\lambda_{1}}{2M^2}\phi_0^4\biggr) RL
\eea
 This contains the conventional radiative correction and renormaliztion group running
 of $\alpha$
 \bea &&
 \alpha \rightarrow \alpha+6\lambda_1( 1-\alpha ) L
 \eea
We  again  use the leading order $R$ equation of motion in the Einstein frame 
as in eq.(\ref{R00}) to implement the contact term, 
but must now keep the $ \lambda^2_1L$ term,
\bea &&
\label{R11}
R
 \rightarrow (\lambda_1 -18\lambda^2_1L) \frac{\phi_0^{4}}{M^2}+{\cal{O}}(\hbar)
\eea
hence
\bea &&
\Gamma_{Jordan}\equiv V_0+\Gamma_0+\Gamma_\alpha+\Gamma_{D3} 
\nonumbo
=
\frac{\lambda _{1}}{4}\phi_0^{4}
+\frac{\lambda _{2}+\alpha\lambda_1}{12}
\frac{\phi_0 ^{6}}{M^{2}}-\frac{2\lambda^2_1\alpha\phi_0^{6}}{M^2}L
\nonumbo
-\half L \biggl(
9\lambda^2_1 \phi_0^{4} +(15{\lambda_1\lambda_2 }+3\lambda^2_1\gamma-[\lambda_1^2])\frac{\phi_0^6}{M^{2}}
\biggr) 
\nonumbo
=
\nonumbo
\frac{\lambda'_{1}}{4}\phi_0^{4}
+\frac{\lambda' _{2}}{12}
\frac{\phi_0 ^{6}}{M^{2}}
+\frac{\lambda'^2_1\alpha\phi_0^{6}}{2M^2}L\biggl(8-{3}\alpha\biggr)
\nonumbo
-\half L \biggl(
9\lambda'^2_1 \phi_0^{4} +(15{\lambda'_1\lambda'_2 }
+3\lambda'^2_1\gamma'-[\lambda'_1{}^2])\frac{\phi_0^6}{M^{2}}
\biggr) 
\nonumbo
\eea
where we've converted to the  Einstein frame primed variables
via eqs.(\ref{stuff0}).
Comparing to eq.(\ref{resultE}) we therefore see an inconsistency between the potentials
 $\Gamma_{Einstein}$ and  $\Gamma_{Jordan}$ at O($\hbar$):
 \bea
 \label{result}
 &&
 \backo
 \Gamma_{Einstein}-\Gamma_{Jordan}=\frac{(8-3\alpha)\alpha\lambda'^2_1\phi_0^{6}}{2M^2}\hbar L 
 \eea
 The fault lies in the Jordan frame calculation  which does not implement the
 contact term.   
 
 One
 might attempt to interpret this as a quantum ``frame anomaly.'' We see some vague parallels with the
 axial anomaly here, but we think the issue is more of an error that one is making in
 computing naively.  Such an anomaly interpretation would require identifying an operator
 that plays a special roll in the Weyl tranformation and that has matrix elements 
 given by the rhs of eq,(\ref{result}).  Perhaps this can be related to the Weyl current,
 but we have not yet explored this possibility.

\section{CONCLUSIONS}

The Weyl transformation acting on the Jordan frame, to remove non-minimal interactions
leading to the minimal Einstein frame, is identical to implementing
the contact terms \cite{HillRoss}.
If one didn't know about the Weyl transformation
one would discover it in the induced contact
terms in the single graviton exchange potential involving non-minimal couplings. 
The Weyl transformation is  powerful as it is fully non-perturbative.
Technically it can provide a useful check on the normalization and implementation
of the graviton propagators in various gauges.  But the contact term stipulates that the
mapping to the Einstein frame is dynamical and inevitable,
and does not involve field redefinitions.

In a model with non-minimal coupling $-\alpha\phi^2R/12$ this implies that the 
parameter $\alpha$ doesn't really exist physically.  Computing $\beta$-functions in a Jordan frame
without implementing the contact term will yield  incorrect results. 
 Implementing the contact term
yields the Einstein frame and results computed there will have no contact term ambiguities.

However, the Einstein frame has a loop
induced contact term which is then absorbed back into the potential terms
by the contact terms (equivalently, a mini-Weyl transformation, 
or use of the $R$ equation of motion). 
The use of the $R$ equation of motion
on the non-minimal term is analogous to the use of the gluon field
equation for the electroweak penguin. It is likely the Deans and Dixon \cite{Deans}
constraints on null operators applies to gravity as well.
This implies generally that there are pitfalls in directly
interpreting any physics in the Jordan frame.  

We emphasize that  {\em  our analysis applies strictly to
a theory with a Planck mass term.}   
A Weyl invariant theory, where $M=0$, is nonperturbative
and our analysis is then inapplicable, and the Jordan frame may then
be physically relevant. Indeed, there is
no conventional gravity in this limit since the usual $M^2R$  (Fierz-Pauli) graviton kinetic 
term does not then exist.  Hence in this limit  one 
would have to appeal to a UV completion, e.g., string theory or $R^2$ gravity, etc.  

In the case of an $R^2$ UV completion theory we  view the formation of
the Planck mass by, e.g., inertial symmetry breaking, i.e., as a dynamical
phase transition, similar to a disorder-order phase transition in
a material medium \cite{FHR}. 
An intriguing point to note is that if we have an $R^2$ UV completion, then the graviton propagator
is $\propto 1/q^4$. But then the $AR\sim Ahq^2$ vertex implies a $1/q^2$ graviton exchange amplitude, which
means an inverse square law ``pseudo-gravitational force'' exists even above the Planck scale
to fields that couple non-minimally.  This is remarkable to us
and one of many issues to develop further in
this context.

\vspace{0.1in}
\noindent
 {\bf Acknowledgements}
\vspace{0.1in}
 
Part of this work was done at 
Fermilab, operated by Fermi Research Alliance, 
LLC under Contract No. DE-AC02-07CH11359 with the United States 
Department of Energy.

\appendix
\section{Projection-Regulated Feynman Loops}

The loop induced effective potential for $\phi_0$ provides a useful way
to extract all of the $\beta$-functions of the various coupling constants.
The potential $\Gamma(\phi_0)$ is the log of the path integral:  $\Gamma=i\ln P$.
In the case of  a real scalar field with mass term
we consider the free action,
\bea &&
\half \int d^4x  \biggl( \partial\phi\partial\phi -m^2\phi^2\biggr)
\eea
we have for the path integral:
\bea
P=\underset{k}{\prod} \biggl(  k^{2}-m^{2} \biggr)^{-1/2}  
=\det \biggl( k^{2}-m^{2} \biggr)^{-1/2}
\eea
whre $k=(k_0, \vec{k})$ is the $4$-momentum
hence, we have:
\bea
\label{A2}
\Gamma=i\ln P= -\frac{i}{2}  \int\frac{d^{4}k}{(  2\pi)  ^{4}}%
\ln \biggl(  k^{2}-m^{2}+i\epsilon \biggr)  
\eea
This can be evaluated with a Wick rotation to a Euclidean momentum, $k\rightarrow k_E=(ik_0,\vec{k})$, and a Euclidean
momentum space cut off $\Lambda$: 
\bea
\label{A3}
\Gamma &= & \frac{1}{2}\int_0^{\Lambda}\frac{d^{4}k_E}{ (  2\pi )^{4}}
\ln \biggl(  \frac{k_E^{2}+m^{2}}{\Lambda^{2}} \biggr) 
\nonumber \\
&&
\backo =
\frac{1}{64\pi^{2}} \biggl( 
\Lambda^{4} \ln\frac{\Lambda^{2}+m^{2}}{\Lambda^{2}}
-m^{4}\ln\frac{\Lambda^{2}+m^{2}}{m^{2}}
\nonumbo
\backo\backo\qquad
-\frac{1}{2} \Lambda^{4}
+\Lambda^{2}m^{2} \biggr)  + (\makebox{irrelevant constants})
\eea
The
cutoff can be viewed is a spurious parameter, introduced to make the integral finite
and arguments of logs dimensionless, but
and not part of the defining
action. 
The only physically meaningful dependence upon $\Lambda$ is contained 
in the
logarithm, where it reflects scale symmetry breaking by the 
quantum trace anomaly. Powers of $\Lambda$, e.g., $\Lambda^{4},\Lambda^{2}m^{2}$, spuriously break scale
symmetry and are not part of the classical action  \cite{Bardeen}.

It is therefore conceptually useful to have a definition of the
loops in which the spurious powers of $\Lambda$ do not arise.
This can be done by defining the loops
applying projection
operators on the integrals.  The projection operator
\bea
P_{n}= \biggl(  1-\frac{\Lambda}{n}\frac{\partial}{\partial\Lambda} \biggr)  
\eea
removes any terms proportional to $\Lambda^{n}.$ 
Since the defining classical Lagrangian
has mass dimension 4 and involves no terms with  $\Lambda^{2}m^{2}$ or
$\Lambda^{4},$ we define the regularized loop integrals as:
\bea
\label{19}
\Gamma &\rightarrow &\frac{1}{2}P_{2}P_{4}\int_0^{\Lambda}\frac{d^{4}k_E}{ (  2\pi )^{4}}
\ln \biggl(  \frac{k_E^{2}+m^{2}}{\Lambda^{2}} \biggr) 
\nonumber \\
&=&
-\frac{1}{64\pi^{2}}  m^{4} \biggl(  \ln\frac{\Lambda^{2}}{m^{2}%
}\biggr)   +O \biggl(  \frac{m^{6}}{\Lambda^{2}} \biggr)    
\eea
where we take the limit $\Lambda>>m$ to suppress $O ( m^{6}/
{\Lambda^{2}})  $ terms and we are interested 
only in the log term (not additive constants)
 This means that the additive, non-log terms, e.g.
$c'm^2$, are undetermined, and the only physically meaningful  result is the 
$\ln(\Lambda^2/m^2)$ term. $\Lambda $ can be swapped for a renormalization scale $\mu$.
Interestingly, if we define the integral as $\int\rightarrow P_1P_2P_3...P_\infty\int$,
the action on the logs will lead to the Euler constant that arises in dimensional regularization,
hinting at a mapping to the dimensionally regularized result.

\end{document}